# DR-Unet104 for Multimodal MRI brain tumor segmentation

Jordan Colman[1,2], Lei Zhang[2], Wenting Duan[2], Xujiong Ye[2]

[1]Ashford and St Peter's Hospitals NHS Foundation Trust, Surrey, UK
jordan.colman@nhs.net,
[2] School of Computer Science, University of Lincoln, Lincoln, UK LZhang@lincoln.ac.uk,
wduan@lincoln.ac.uk, XYe@lincoln.ac.uk

**Abstract.** In this paper we propose a 2D deep residual Unet with 104 convolutional layers (DR-Unet104) for lesion segmentation in brain MRIs. We make multiple additions to the Unet architecture, including adding the 'bottleneck' residual block to the Unet encoder and adding dropout after each convolution block stack. We verified the effect of including the regularization of dropout with small rate (e.g. 0.2) on the architecture, and found a dropout of 0.2 improved the overall performance compared to no dropout, or a dropout of 0.5. We evaluated the proposed architecture as part of the Multimodal Brain Tumor Segmentation (BraTS) 2020 Challenge and compared our method to DeepLabV3+ with a ResNet-V2-152 backbone. We found the DR-Unet104 achieved a mean dice score coefficient of 0.8862, 0.6756 and 0.6721 for validation data, whole tumor, enhancing tumor and tumor core respectively, an overall improvement on 0.8770, 0.65242 and 0.68134 achieved by DeepLabV3+. Our method produced a final mean DSC of 0.8673, 0.7514 and 0.7983 on whole tumor, enhancing tumor and tumor core on the challenge's testing data. We produce a competitive lesion segmentation architecture, despite only using 2D convolutions, having the added benefit that it can be used on lower power computers than a 3D architecture. The source code and trained model for this work is openly available at https://github.com/jordan-colman/DR-Unet104.

# 1      Introduction

Lesion segmentation is an important area of research necessary to progress the field of radiomics, using imaging to infer biomarkers, that can be used to aid prognosis prediction and treatment of patients [1]. Segmentation of gliomas, the most common form of primary brain malignancy [2], is a highly useful application of lesion segmentation. Accurate brain tumor segmentation in MRI produces useful volumetric information, and in the future may be used to derive biomarkers to grade gliomas and predict prognosis. Manual brain tumor segmentation, however, is a skilled and time-consuming task. Therefore, automated brain tumor segmentation would be of great benefit to progress this area. However, accurate segmentation remains a challenging task due to the high variation of brain tumor size, shape, position and inconsistent intensity and contrast in various image modalities. This has contributed to the development of automatic segmentation methods, for which several methods have been proposed. The Multimodal Brain Tumor Segmentation (BraTS) challenge is an annual challenge set to act as a benchmark of the current state-of-the-art brain tumor segmentation algorithms.

In the recent years deep neural networks have achieved the top performance in the BraTS challenge. Many existing methods consider the Unet [3] as a base architecture, a basic but effective form of the encoder-decoder network design, with at least two winners of the BraTS challenge utilising a variation of the Unet in 2017 and 2019 [4,5]. Other current state-of-the-art segmentation algorithms use the ResNet [6,7] as an encoder in such architecture, such as the DeepLabV3+, which uses ResNet-101 as the initial encoder and spatial pyramid pooling module in the final layer [8]. The ResNet uses identity mapping, a residual connection which skips multiple network layers to aid back propagation and allows deeper networks to be made. Additionally, the ResNet uses a 'bottleneck' residual block which uses a 1x1 convolution to reduce the number of image features prior to a spatial, or 3x3 convolution, and then uses another 1x1 convolution to increase the number of features. This is done to increase computational efficiency and at the same time increase the number of image features represented in the network. The ResNet and DeeplabV3+ do not use random dropout, a commonly used regularizer, which randomly removes the signal of a given proportion of neurons in a layer in order to reduce overfitting of training data [9].

A common approach of improving the performance of current architectures in medical image segmentation is by extending 2D image segmentation to 3D, using 3D convolutional networks on whole images as opposed to a single 2D MRI slice. An example is the BraTS 2019 1$^{st}$ place solution for the brain tumor segmentation task, which used a two-staged cascaded 3D Unet [4]. The paper used a 3D Unet architecture with the addition of residual connections in convolutional blocks, in stacks of 1, 2 or 4. The first cascade has fewer total feature channels and detects 'coarse' features, the second cascade detects finer details as it has more total feature channels. This architecture achieved a mean dice score coefficient (DSC) of 0.8880 for whole tumor segmentation of the testing data, 0.8370 for tumor core and 0.8327 for enhancing tumor. However, this architecture required cropping 3D images to run in a batch size of 1 on a graphics card with 12Gb of memory due to the high memory usages of 3D processing and the large image size.

In this paper, given the success of the Unet in BraST challenge [4,5] and inspired by the 'bottleneck' residual block from the ResNet [6,7] we proposed a new architecture to couple the strengths of the 'bottleneck' residual block and the Unet for brain tumor segmentation in the BraTS 2020 challenge. The proposed network has a total of 104 convolutional layers, so is named, deep residual Unet 104 (DR-Unet104). We additionally include dropout and investigate if this improves architecture performance, as we mimic the ResNet in our encoder and it is suggested by its creators additional regularization may improve performance [7].

## 2 Methods

### 2.1 Architecture

The proposed architecture DR-Unet104 overview is shown below in Figure 1 and 2. The detailed architecture description is shown in Figure 3, in order to display the number of image feature channels in each layer. It comprises of three main components: encoder, decoder and bridge that forms a typical U-shape network. In this design, five stacked residual block levels with convolution layers and identify mapping are deployed in both encoder and decoder components, which are connected by the bridge component. The feature representations are encoded by the encoder path, which are recovered in the decoder path to a pixel-wise classification. Figure 2 shows the outline of the residual blocks in the encoder and decoder path. In the encoder path and bridge connection, the bottleneck design is applied, which consists of a 1×1 convolution layer for reducing the depth of feature channel, a 3×3 convolution layer, and then a 1×1 convolution layer for restoring dimension [6]. In the decoder path, the typical residual block consists of two stacked 3x3 2D convolutions. The batch normalisation (BN) and rectified linear unit (ReLU) activation are used in all residual blocks (Figure 2), we use 'pre-activation' residual connections as used in ResNet-V2 and described in He *et al.* 2016 [7].

Given a deep architecture has many layers, the issues regarding overfitting and dead neurons in activation need to be considered for training the network. In our method, following the work [9], we employ a regularization using dropout with a small rate (e.g. dropout rate of 0.2) after each level. In our method, the input of each level (after upsampling) of the decoder is added with a concatenation connection from the output of the encoder to aid feature mapping. Downsampling is performed with a stride of 2 in the first convolution of each level (except for level 1). Upsampling is performed with 2D transposed convolution with a kernel of 2x2 and stride of 2. The final layer is convoluted with a 1x1 kernel and generating pixel-wise classification scores to represent the 3 tumor classes, and background class, the class of the pixel is decided by the channel with the largest output (argmax), softmax is used during training. The input is a 2D image slice with 4 channels representing each MRI modality. The proposed network code is publicly available at https://github.com/jordan-colman/DR-Unet104.

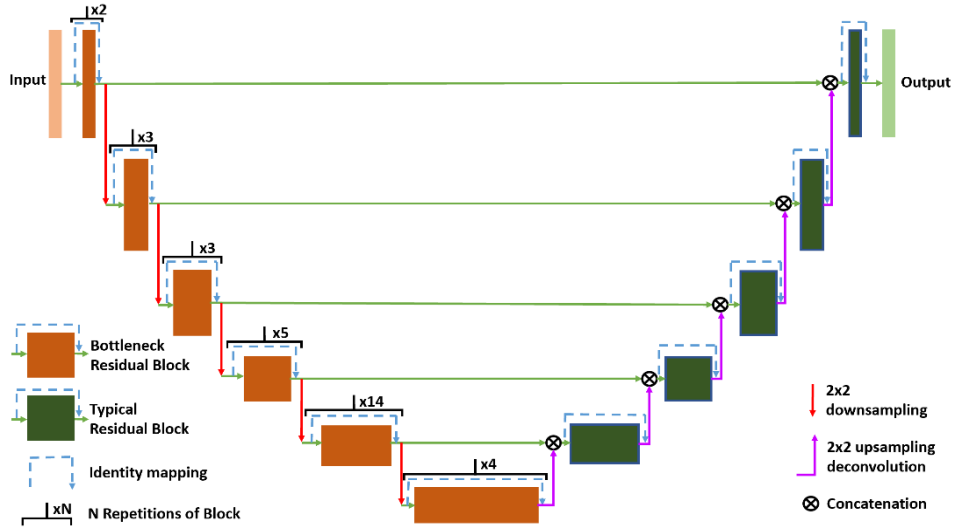

**Figure 1.** Overview of the DR-Unet104 architecture showing the bottleneck residual block in the encoder and typical residual block in the decoder. The number of stacks of the bottleneck block is also shown. The downsampling is performed by a step of 2 in the initial 1x1 convolution in the first bottleneck block of the level.

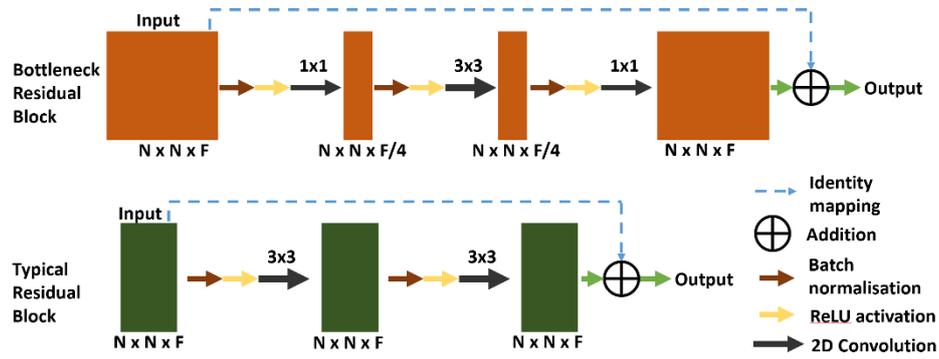

**Figure 2.** The outline of the typical residual blocks in the decoder and the bottleneck residual block of the encoder path. The typical residual block can be seen to be formed of two 3x3 2D convolutions with batch normalization and rectified linear unit (Relu) activation before each convolution. The Bottleneck residual block has a 1x1 2D convolution, which reduces the number of image feature channels (F) to ¼ of the number. This is followed by a 3x3 2D convolution then a 1x1 convolution which increases the feature channel number by 4 times to the input number. Both blocks input is added to the output for the identity mapping to aid backpropagation [6,7].

| | Image dimensions | Encoder | Decoder |
|---|---|---|---|
| **DR-Unet104** | | | |
| **Input / output & Level 1** | 240 x 240 | Input $\begin{bmatrix} 1x1, 16 \\ 3x3, 16 \\ 1x1, 64 \end{bmatrix} x2$ (Dropout 0.2) | Output Argmax $1 \times 1, n\_class$ $\begin{bmatrix} 3 \times 3, 32 \\ 3 \times 3, 32 \end{bmatrix} x1$ → ↑(Dropout 0.2) |
| **Level 2** | 120 x 120 | ↓ $\begin{bmatrix} 1x1, 32 \\ 3x3, 32 \\ 1x1, 128 \end{bmatrix} x3$ (Dropout 0.2) | $\begin{bmatrix} 3 \times 3, 64 \\ 3 \times 3, 64 \end{bmatrix} x1$ → ↑(Dropout 0.2) |
| **Level 3** | 60 x 60 | ↓ $\begin{bmatrix} 1x1, 64 \\ 3x3, 64 \\ 1x1, 256 \end{bmatrix} x3$ (Dropout 0.2) | $\begin{bmatrix} 3 \times 3, 128 \\ 3 \times 3, 128 \end{bmatrix} x1$ → ↑(Dropout 0.2) |
| **Level 4** | 30 x 30 | ↓ $\begin{bmatrix} 1x1, 128 \\ 3x3, 128 \\ 1x1, 512 \end{bmatrix} x5$ (Dropout 0.2) | $\begin{bmatrix} 3 \times 3, 256 \\ 3 \times 3, 256 \end{bmatrix} x1$ → ↑(Dropout 0.2) |
| **Level 5** | 17 x 17 | ↓ $\begin{bmatrix} 1x1, 256 \\ 3x3, 256 \\ 1x1, 1024 \end{bmatrix} x14$ (Dropout 0.2) | $\begin{bmatrix} 3 \times 3, 512 \\ 3 \times 3, 512 \end{bmatrix} x1$ → ↑(Dropout 0.2) |
| **Bridge** | 8 x 8 | ↓ $\begin{bmatrix} 1x1, 512 \\ 3x3, 512 \\ 1x1, 2048 \end{bmatrix} x4$ | |

**Figure 3.** Our Proposed architecture for the DR-Unet104. The [] brackets denote the typical or bottleneck residual block with 3x3,64 representing a 2D convolution with a 3x3 kernel and 64 layers. $xN$ denotes the number, N, of stacked residual blocks in that layer. ↓ denotes reduction in spatial resolution performed by a stride of 2 in the first 1x1 2D convolution of the initial bottleneck residual block of that level. ↑ denotes upsampling via the 2D transposed convolution with a kernel of 3x3 and stride of 2. → denotes a skip connection between the output of the encoder with the input of the decoder at the same level, joined via concatenation to the unsampled output of the previous level.

## 2.2   Loss function

The loss function used was sparse categorical cross entropy (CE), which is calculated with Eq (1). This was chosen for simplicity, however, better performance may have been produced using other loss functions such as 'soft Dice loss' [4].

$$CE = -\frac{1}{N}\sum_n^N \sum_c^C Y_{true_c}^n \times \log\left(Y_{pred_c}^n\right) \qquad (1)$$

Where $N$ is number of examples and $C$ represents the classes, $Y_{true}$ is the truth label and $Y_{pred}$ the softmax of the prediction [10].

## 3    Experiment

### 3.1    Dataset, Pre-processing and data augmentation

The data used for training and evaluation of our model consisted of the BraTS 2020 dataset including training data with 369 subjects, validation data with 125 subjects and testing data with 166 subjects, all contained low and high grade gliomas [11-15]. Each subject has a T2 weighted FLAIR, T1 weighted, T1 weighted post contrast, and a T2 weighted MRI. Each image is interpolated to a 1x1x1 mm voxel sized giving images sized 240x240x155 voxels. All subject images are aligned into a common space and skull stripped prior to data sharing. The training data additionally contains manually drawn tumor segmentation mask, annotated by experts and checked by a neuroradiologist, the segmentation is labelled with numbers denoting edema, tumor core and enhancing tumor. Figure 4 shows an example of a subjects imaging modalities and tumor mask.

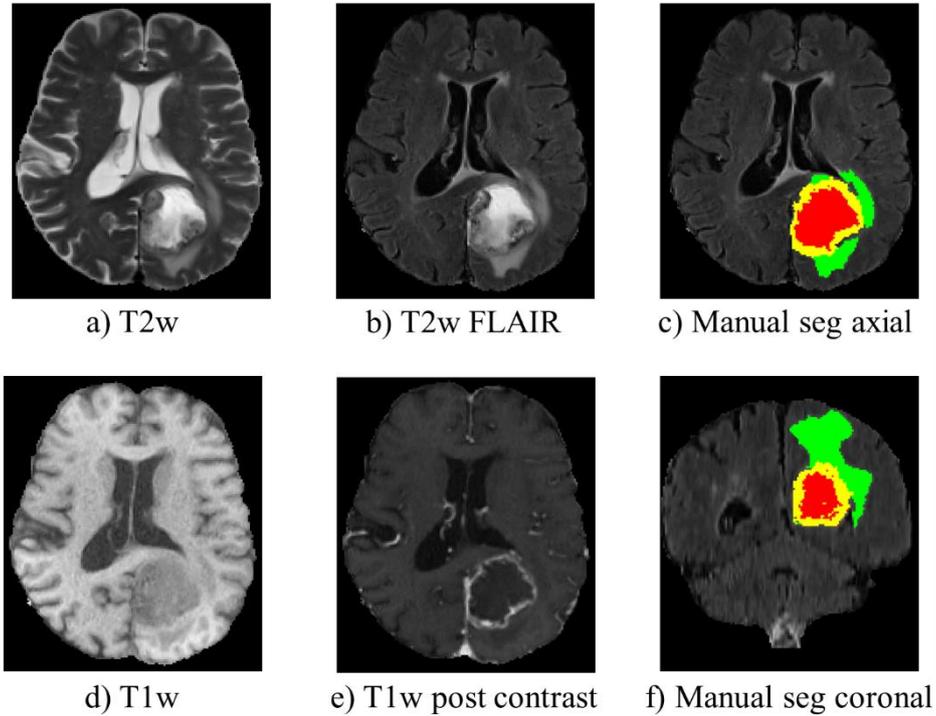

**Figure 4.** Shows the different MRI modalities provided on one subject with a glioma, a) shows a T2 weighted MRI, b) a T2 weighted FLAIR MRI, d) a T1 weighted MRI, e) a T1 weighted MRI post contrast. c) and f) show the manual tumor segmentation in an axial and coronal slice respectively on a T2 weighted FLAIR MRI. Green is edema, red tumor core and yellow enhancing tumor.

We pre-processed the images using data standardisation on a whole 3D MRI modality wise basis, normalised using Eq (2) where $v$ is a given voxel and the $mean$ and standard deviation ($SD$) are of the image of voxels $> 0$. This is so the image was rescaled prior to conversion to 2D image.

$$v = \begin{Bmatrix} 254 & v > mean + 3*SD \\ \frac{(v-mean)+127}{3*SD/128} & otherwise \\ 0 & v < mean - 3*SD \end{Bmatrix} \qquad (2)$$

Each slice of the image is then saved as individual png image, with each channel representing an MRI modality. The only image augmentation applied was randomly flipping the images in the left-right and anterior-posterior orientation with a 50% probability.

### 3.2   Setup and training

For training, we used ADAM optimiser [16] with a learning rate of 1e-4 and He initialization [17]. We used a batch sizes of 10 and ran the training for 50 epochs. All subjects' images were converted to single RGBA images for each slice as described in the methods and inputted to the model in a random order. We evaluated our method with varying dropout rates and compared our proposed architecture to DeeplabV3+ as the architecture is openly available online and a current state-of-the-art segmentation architecture. We use a ResNet-V2-152 backbone for the BraTS evaluation and not the originally suggested ResNet-V1-101 in order to improve the performance and make the architecture more analogous to our proposed method [6]. The architectures are implemented in Keras with a tensorflow backend [18-19]. This was performed using a NVIDIA GTX 2080Ti graphics card with 12 GB of memory.

## 4   Results

We initially evaluated our proposed DR-Unet104 architecture on the 125 validation subjects, pre-processed in the same way as the training data, however, without data augmentation. The resulting 2D masks are reconstructed into 3D nii images and evaluated on the IPP website [https://ipp.cbica.upenn.edu/], which computed and evaluated the lesion masks outputting the metrics, DSC, sensitivity, specificity and Hausdorff distance 95% (HD95). We evaluated our proposed architecture with varying random dropout rate after each level, the rate we trailed being 0.2, 0.5 and no dropout. We can observe from the Table 1, the architecture applying the dropout of rate 0.2 has superior performance than the architecture with a rate of 0.5 or without dropout, with all tumor components' DSC and HD95 being greater. These observations verify that dropout in our architecture allows the network to learn informative features from the data, but only with the small rate (e.g. 0.2), as the network with a large (0.5) rate performs poor and even worse than the setting without dropout. Comparing our proposed architecture with a droupout of 0.2 to DeepLabV3+ shows improvement of segmentation results in all areas except the mean tumor core DSC.

|  | Validation Results | | | | | |
|---|---|---|---|---|---|---|
|  | DSC | | | HD95 | | |
| Architecture | WT (SD) | ET (SD) | TC (SD) | WT (SD) | ET (SD) | TC (SD) |
| DR-UNET104 dropout (0) | 0.8763 (0.0859) | 0.6549 (0.3256) | 0.6693 (0.3357) | 18.39 (24.87) | 53.61 (115.9) | 16.19 (20.88) |
| DR-UNET104 dropout (0.2) | **0.8862** **(0.0886)** | **0.6756** **(0.3171)** | 0.6721 (0.3462) | **12.11** **(20.82)** | **47.62** **(112.7)** | **15.74** **(36.04)** |
| DR-UNET104 dropout (0.5) | 0.8723 (0.0977) | 0.6701 (0.3245) | 0.6489 (0.3651) | 23.80 (27.58) | 51.53 (108.5) | 28.56 (51.69) |
| DeepLabV3+ | 0.8771 (0.0853) | 0.6524 (0.3101) | **0.6813** **(0.3213)** | 14.87 (23.20) | 49.10 (112.6) | 17.96 (37.54) |

**Table 1.** Table showing the mean results of the Dice Score coefficient (DSC) and Hausdorff distance 95% (HD95) with the Standard deviation (SD) in brackets results on the BraTS20 validation data. We show results for the whole tumor (WT), enhancing tumor (ET) and tumor core (TC). We compare the results of our own architecture, the deep residual Unet 104 (DR-Unet104) with no dropout, a dropout of 0.2 and of 0.5 after each level. We additionally compare our model to DeepLabV3+ with a ResNet101 backbone. We show the best results in bold.

We finally evaluated our model on the 166 BraTS testing subjects as part of the 2020 challenge with the final results shown in Table 2. Our mean whole tumor DSC is lower than the validation results at 0.8673, however the enhancing tumor and tumor core DSC are much higher than validation results achieving 0.7514 and 0.7983 respectively. The proposed methods overall better performance on the testing data shows it generalizability aided by random dropout. Our model's performance is additionally competitive with 3D models.

|  | Testing Data | | | | | |
|---|---|---|---|---|---|---|
|  | DSC | | | HD95 | | |
| Architecture | WT (SD) | ET (SD) | TC (SD) | WT (SD) | ET (SD) | TC (SD) |
| DR-UNET104 dropout (0.2) | 0.8673 (0.1279) | 0.7514 (0.2478) | 0.7983 (0.2757) | 10.41 (16.59) | 24.68 (83.99) | 21.84 (74.35) |

**Table 2.** Table showing the mean results of the Dice Score coefficient (DSC) and Hausdorff distance 95% (HD95) with the Standard deviation (SD) in brackets results on the BraTS20 Testing data. We show results for the whole tumor (WT), enhancing tumor (ET) and tumor core (TC) evaluated using the deep residual Unet 104 (DR-Unet104) with dropout of 0.2.

## 5 Discussion

In this paper, we present a 2D deep residual Unet with 104 convolution layers for automated brain tumor segmentation in multimodal MRI. The proposed network couples the strengths of deep residual blocks and the Unet with encoder-decoder structure. The regularization of dropout is included into the network, allowing it to learn more representative features than the plain architectures without regularization, producing improved validation results as shown in table 1.

The results show that our method achieves promising performance when comparing to a state-of-the-art method (i.e. DeeplabV3+), and performs reasonably well despite being a 2D architecture, having minimal training data augmentation and being trained for only 50 epochs without any complex learning rate scheduling. The 2D architecture has the added benefit of meaning the model can be evaluated on lower powered computers/GPUs, only needing a GPU with 1-2 GBs of memory (when evaluated with a batch size of 1), unlike many other 3D architectures [4]. A limitation is that for simplicity we used the commonly used cross-entropy loss function, but would likely have received a better performance using a 'soft Dice loss' function, as used by the BraTS 2019 1st place method [4], due to DSC being used for the evaluation, and this could be included in future work.

Unusually, on the testing data set our architecture performed worse on whole tumor segmentation, but much better on enhancing tumor and tumor core labelling, compared to the validation data set. This is likely due a greater number of difficult to segment enhancing areas in the validation data set, which would also affect the tumor core DSC if mislabelled as one another. A greater number of difficult to label enhancing tumor areas in the validation data set is supported by a larger standard deviation of the mean DSC, 0.25 in testing vs. 0.32 on validation data.

## 6 Conclusion

We propose a variant of the Unet taking advantage of bottleneck residual block to produce a deeply stacked encoder. We additionally show the benefit of using dropout in our architecture. Our method has a competitive performance despite being a 2D architecture and having simple and limited training.